\pdfoutput=1
\documentclass[%
 reprint,
amsmath,amssymb,
aps,
prl
]{revtex4-1}

\usepackage{graphicx}% Include figure files
\usepackage{dcolumn}% Align table columns on decimal point
\usepackage{bm}% bold math

\usepackage{amsmath,amssymb,amsbsy,amstext,amsthm,simplewick,amsfonts,graphicx}
\usepackage{mathrsfs}
\usepackage{graphicx}
\usepackage{wrapfig}
\usepackage{upgreek}
\usepackage{bm} %-> bold math
\usepackage{framed}
\usepackage{bbm}
\usepackage{textcomp}
\usepackage{adjustbox}
\usepackage{makecell}
\usepackage{tcolorbox}
\usepackage{empheq}
\usepackage[normalem]{ulem}
\usepackage{enumitem}
\usepackage{braket} 
\usepackage{array}
\usepackage{dsfont}
\usepackage{ulem}
\usepackage{xcolor}
\usepackage{caption}
\usepackage{subcaption}
\usepackage{tocloft}
\usepackage{tikz}
\usepackage{xcolor}
\usepackage{leftindex}
\usetikzlibrary{patterns}
\usetikzlibrary{patterns}
\usetikzlibrary{decorations.markings}
\usetikzlibrary{decorations.pathmorphing}
\usetikzlibrary{shapes.misc}
\usetikzlibrary{arrows}
\usetikzlibrary{shapes.geometric}
\usetikzlibrary{plotmarks}
\usepackage{hyperref}
\hypersetup{colorlinks=true,
	breaklinks=true,
	pdfstartview=Fit,
	linkcolor=blue2,
	citecolor=red2,
	urlcolor=blue
}

\definecolor{red2}{RGB}{214, 39, 40}
\definecolor{green2}{RGB}{0,170,0}
\definecolor{blue2}{RGB}{0,100,200}
\definecolor{magenta2}{RGB}{191,64,191}
\definecolor{purple2}{RGB}{112,48,160}
\definecolor{orange2}{RGB}{255,192,0}

\usepackage{graphicx,scalerel}

\graphicspath{{Fig/}}

\def\Re{\mathrm{Re}\,}
\def\Im{\mathrm{Im}\,}
\newtheorem{conjecture}{Conjecture}

\definecolor{rosewood}{rgb}{0.4, 0.0, 0.04}
\definecolor{pyblue}{RGB}{31, 119, 180}
\definecolor{pyorange}{RGB}{255, 127, 14}
\definecolor{pygreen}{RGB}{44, 160, 44}
\definecolor{pyred}{RGB}{214, 39, 40}
\definecolor{pypurple}{RGB}{148, 103, 189}
\definecolor{lightgray}{gray}{0.9}

\begin{document}

\preprint{APS/123-QED}

\title{New Exceptional EFTs in de Sitter Space
from \\
Generalised Energy Conservation}

\author{Zong-Zhe Du}
\email{zongzhe.du@nottingham.ac.uk}
\affiliation{
School of Physics and Astronomy,
			University of Nottingham, University Park, \\ Nottingham, NG7 2RD, UK}

\author{David Stefanyszyn}
\email{david.stefanyszyn@nottingham.ac.uk}
\affiliation{
School of Mathematical Sciences \& School of Physics and Astronomy, University of Nottingham,\\ University Park, Nottingham, NG7 2RD, UK}

\date{\today}

\vspace{20mm}

\begin{abstract}
\vspace{10mm}

We discover a surprising relationship between exceptional effective field theories (EFTs) in de Sitter space and a notion of \textit{generalised energy conservation} (GEC) of an S-matrix defined in an extended Poincar\'{e} patch of four-dimensional de Sitter. By demanding that such an S-matrix only has support when the total energies of in and out states are equal, we constrain the coupling constants in theories of self-interacting scalars living in the exceptional series of de Sitter representations. We rediscover the theories of Dirac-Born-Infeld (DBI) and Special Galileon, and when increasing the conformal dimension we find evidence for new exceptional theories where the four-point scalar self interactions are uniquely fixed in terms of a single coupling constant. We conjecture that for each integer conformal dimension $\Delta \geq 4$, there is at least one exceptional EFT that can be entirely fixed by GEC.
\end{abstract}

\maketitle

\newpage

\newpage

\section{Introduction and Summary}
Energy conservation is usually perceived as a consequence of global time-translation symmetry by Noether's theorem \cite{Noether:1918zz}. In flat space where time translation is an exact symmetry, the probability for the incoming and outgoing states to have different total energy is exactly zero. When there is no time translation symmetry as in de Sitter (dS) space, however, energy conservation should not be imposed a priori even if we have asymptotic states where well-separated wave packets can scatter.

Recently, the authors of \cite{Donath:2024utn} suggested that one can extend the Poincar\'{e} patch of dS space by (we will discuss the $\eta=0$ elephant in the room in detail in this work):
\begin{align}
    ds^2 = \frac{-d\eta^2 + d\vec{x}^2}{H^2 \eta^2}, \;\;\eta\in(-\infty,+\infty)\,. \label{Metric}
\end{align}
 The idea is to glue together two Poincar\'{e} patches of dS in order to define an in-out correlator from which one can define an S-matrix via an LSZ reduction (see \cite{Melville:2023kgd,Melville:2024ove,Marolf:2012kh,Albrychiewicz:2020ruh} for other definitions of scattering in dS). The $\alpha \rightarrow \beta$ element of this S-matrix can be defined in the interaction picture by the Dyson series \cite{Donath:2024utn}  
\begin{align}
    S_{\beta\alpha} &\equiv  \braket{\beta|\mathcal{T}\{e^{-i\int_{-\infty}^{+\infty} d\eta\mathcal{H}_{\text{int}} }\}|\alpha} \\
    & \equiv 1 + i (2\pi)^4  \mathcal{A} ~ \delta^{3}(\vec{k}_{\text{in}}-\vec{k}_{\text{out}}) \label{DysonSeries}\,,
\end{align}
where $\mathcal{H}_{\text{int}}$ is the interaction Hamiltonian in the interaction picture, and $\beta,\;\alpha$ are out and in states in the interaction picture, respectively. The delta function enforcing momentum conservation $\delta^{3}(\vec{k}_{\text{in}}-\vec{k}_{\text{out}})$ arises as a consequence of the spatial translation symmetry of \eqref{Metric}. 

In \cite{Donath:2024utn} it was suggested that this object should be considered for theories that are \textit{manifestly} infrared (IR)-finite where there is no divergence in the interacting Lagrangian as $\eta \rightarrow 0$. In such cases, every contribution to the S-matrix contains an energy-conserving delta function, or derivatives thereof, thereby ensuring a generalised version of energy conservation:
\begin{align}
\mathcal{A}  = \sum_{m=0} \mathcal{A}^{(m)} \partial_{k_{T}}^{m} \delta (k_{T}) \,,
\end{align}
where $k_{T} \equiv \sum_{a\in \text{in}}k_a - \sum_{b\in\text{out}}k_b$, $k_a$ are the magnitudes of three-momenta and the amplitudes $\mathcal{A}^{(m)}$ are functions of the external kinematics. Such energy-conserving delta functions, when scattering e.g. conformally-coupled or exceptional series scalars, arise from integrals of the form $\int_{- \infty}^{+ \infty} d \eta \eta^{p} e^{-i k_{T} \eta}$ with integer $p \geq 0$ \cite{Donath:2024utn}. 

In this work, however, we allow ourselves to consider theories that are \textit{not} manifestly IR-finite. By imposing Bunch-Davies (BD) vacuum conditions for both the asymptotic past and future, where the free vacuum $\ket{0}$ ($\bra{0}$) is annihilated by the annihilation (creation) operator $a_{\vec{k}}$ ($a_{\vec{k}}^\dag$), we consider scattering processes by shooting particles from the far past to the far future with a contour deformation around $\eta=0$. In our case, an extra term $\mathcal A_{k_T\neq 0}$ is in general possible in the S-matrix without any energy-conserving delta function:
\begin{align}
\mathcal{A} = \mathcal{A}^{(\pm)}_{k_{T} \neq 0} +   \sum_{m=0} \mathcal{A}^{(m)} \partial_{k_{T}}^{m} \delta (k_{T}) \,.
\end{align}
Such an amplitude therefore has support when energy is not conserved. However, there are perhaps good reasons to demand $\mathcal{A}^{(\pm)}_{k_{T} \neq 0} = 0$ and therefore \textit{generalised energy conservation} (GEC). Indeed, a non-zero $\mathcal{A}_{k_{T}\neq 0}^{(\pm)}$ implies eternal energy creation or annihilation, which could lead to instabilities. 

The calculation of $\mathcal{A}_{k_T \neq 0}^{(\pm)}$, and indeed the full amplitude, involves choosing between two different time integration contours $(\pm)$, as shown in Figures \ref{EnergyAnnihilation} and \ref{EnergyCreation}. For $k_T \leq 0$ ($k_T \geq 0$), the contour in Figure \ref{EnergyAnnihilation} (Figure \ref{EnergyCreation}) encloses the potential pole at the origin, whereas for the contour in Figure \ref{EnergyCreation} (Figure \ref{EnergyAnnihilation}) it does not. In all cases the arc at infinity vanishes by our BD boundary conditions, and so the integral along the real line (which computes the amplitude) can be computed by residue theorem. Only the residue at $\eta=0$ can contribute to $\mathcal{A}^{(\pm)}_{k_{T} \neq 0}$ since all other non-zero contributions come from integrals of the form $\int_{- \infty}^{+ \infty} d \eta \eta^{p} e^{-i k_{T} \eta}$ with integer $p \geq 0$ which yield energy-conserving contributions only, as above (we go through this in detail in the Supplementary Material \footnote{See Supplementary Material at [url] for details on how to compute the amplitude}, which includes the reference \cite{Gell-Mann:1951ooy}). Imposing the absence of instabilities is therefore equivalent to imposing that the residue at $\eta=0$ vanishes. Surprisingly, we find that such a condition can impose relations between various coupling constants appearing in general Lagrangians that are not manifestly IR-finite. Imposing GEC can therefore be used to pick out special theories. 

We concentrate on massive scalars that live in the exceptional series of four-dimensional de Sitter representations with integer conformal dimension $\Delta \geq 3$, and find that by imposing GEC there is a unique four-point amplitude for each mass where $m^2 = \Delta (3 - \Delta)H^2$, when we allow for operators with at most $2 \Delta-4$ derivatives. We focus on these masses since such scalars in dS space are naturally packaged with non-linearly realised symmetries \cite{Bonifacio:2018zex}. Despite the tachyonic massess, these scalars are indeed unitary irreducible representations of dS isometries (see e.g. \cite{Sun:2021thf,Gazeau:2010mn,Penedones:2023uqc} for discussions). We recover the known theories of Dirac-Born-Infeld (DBI) ($\Delta = 4, m^2 = -4 H^2$), Special Galileon ($\Delta = 5, m^2 = -10H^2$) for the first two tachyonic exceptional series masses, and find evidence for the existence of new theories beyond these known examples as we continue to increase the conformal dimension of the field. In the cases we consider, we find that the four-point interactions are completely fixed in terms of a single coupling constant. We therefore refer to such theories as \textit{exceptional EFTs}. We will be primarily interested in four spacetime dimensions. However, in the supplementary material we extend our analysis to other dimensions and find that our techniques are equally powerful there too. 

The space of exceptional EFTs in dS space for single-scalar theories has been considered from algebraic and brane-world constructions \cite{Burrage:2011bt,Bonifacio:2018zex,Bonifacio:2021mrf, Garoffolo:2025igz,Goon:2011uw,Goon:2011qf} and wavefunction soft limits \cite{Armstrong:2022vgl} (see also \cite{Bittermann:2022nfh}). No evidence of new single-scalar theories beyond the Special Galileon has been found. These results suggest that the new theories we have discovered in this work might only exist when coupled to additional degrees of freedom. There is evidence that such theories do exist in dS given the existence of certain extensions of the dS algebra \cite{Joung:2015jza}. It would be very interesting to find the symmetries/algebras and spectra that define the theories we have found evidence for in this paper. Furthermore, the actions for such special theories in dS are rather complicated and call for a more ``on-shell" approach to understanding their properties. We hope that our work initiates such a study.

\section{Notation}
We use $k \equiv |\vec{k}|$ for momentum magnitudes, and will often write $x \equiv (\eta, \vec{x})$ with the spacetime index suppressed. We use $c_{0}^{(\Delta)}$ for cubic couplings and $d_{2m}^{(\Delta)}$ for quartic couplings with $2m$ derivatives. After imposing momentum conservation we can express four-point amplitudes in terms of the following six variables:
\begin{align}
k_1, \quad  k_2, \quad k_3,\quad k_4,\quad \vec{k}_1 \cdot \vec{k}_2,\quad \vec{k}_1 \cdot \vec{k}_3. 
\end{align}
There may be further redundancies forced by energy-conserving delta functions.

\begin{figure}
    \centering
\begin{center}
    \centering
    \begin{tikzpicture}[scale = 2]
        \draw[black, ->] (-1.6,0) -- (1.6,0) coordinate (xaxis);
        \draw[black, ->] (0,-1.6) -- (0,1.7) coordinate (yaxis);
        \node at (1.85, 0) {$\Re\eta$};
        \node at (0, 1.85) {$\Im\eta$};
		
        \draw[pyred, fill = pyred] (0,0) circle (.03cm);
        
        \node at (-0.8, -0.25) {\textcolor{pypurple}{$\mathcal C_1$}};
        \node at (0.8, -0.25) {\textcolor{pypurple}{$\mathcal C_3$}};
        \node at (-0.24, 0.24) {\textcolor{pypurple}{$\mathcal C_2$}};
        \node at (1.2, -1.2) {\textcolor{pyblue}{$\mathcal C_4$}};
        \node at (1.2, 1.3) {\textcolor{pyred}{$\mathcal C_4'$}};

        \path[pyred, draw, line width = 1.0pt] (0.25,0.55) -- (0.6,0.55);
        \path[pyblue, draw, line width = 1.0pt] (0.25,-0.55) -- (0.6,-0.55);
        \node at (0.9, 0.55) {$k_T\leq 0$};
        \node at (0.9, -0.55) {$k_T\geq 0$};

        \path[pypurple, draw, line width = 0.8pt, postaction = decorate, 
		decoration={markings,
			mark=at position 0.5 with {\arrow[line width=1pt]{>}}}] (-1.5, 0) -- (-0.15,0);
            \path[pypurple, draw, line width = 0.8pt, postaction = decorate, 
		decoration={markings,
			mark=at position 0.5 with {\arrow[line width=1pt]{>}}}] (0.15,0) -- (1.5,0);
	\path[pyred, draw, dashed, line width = 0.8pt, postaction = decorate, 
		decoration={markings,
			mark=at position 0.25 with {\arrow[line width=1pt]{>}}}] (1.5,0.) arc (-3.81407:180+3.81407:1.50333);

	\path[pyblue, draw, dashed, line width = 0.8pt, postaction = decorate, 
		decoration={markings,
			mark=at position 0.25 with {\arrow[line width=1pt]{>}}}] (1.5,0) arc (-3.81407:-180+3.81407:1.50333);
	\path[pypurple, draw,  line width = 0.8pt, postaction = decorate, 
		decoration={markings,
			mark=at position 0.25 with {\arrow[line width=1pt]{>}}}] (-0.15,0) arc (-180:0:0.15);
	\end{tikzpicture}
\end{center}
    \caption{The time integration contour for a dS scattering amplitude in an eternal energy-creating universe. The contour is chosen such that the integral on the large arc vanishes for $k_T\neq 0$ while the scattering amplitude is non-vanishing only when $k_T = k_{\text{in}} - k_{\text{out}}\leq 0$.}
    \label{EnergyAnnihilation}
\end{figure}
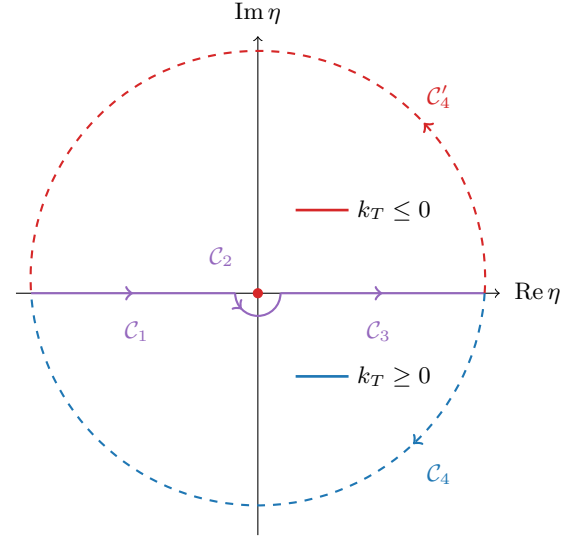

\begin{figure}
    \centering
\begin{center}
    \centering
    \begin{tikzpicture}[scale = 2]
        \draw[black, ->] (-1.6,0) -- (1.6,0) coordinate (xaxis);
        \draw[black, ->] (0,-1.6) -- (0,1.7) coordinate (yaxis);
        \node at (1.85, 0) {$\Re\eta$};
        \node at (0, 1.85) {$\Im\eta$};
		
        \draw[pyred, fill = pyred] (0,0) circle (.03cm);
        
        \node at (-0.8, -0.25) {\textcolor{pypurple}{$\mathcal C_1$}};
        \node at (0.8, -0.25) {\textcolor{pypurple}{$\mathcal C_3$}};
        \node at (-0.24, 0.24) {\textcolor{pypurple}{$\mathcal C_2$}};
        \node at (1.2, -1.2) {\textcolor{pyblue}{$\mathcal C_4$}};
        \node at (1.2, 1.3) {\textcolor{pyred}{$\mathcal C_4'$}};

        \path[pyred, draw, line width = 1.0pt] (0.25,0.55) -- (0.6,0.55);
        \path[pyblue, draw, line width = 1.0pt] (0.25,-0.55) -- (0.6,-0.55);
        \node at (0.9, 0.55) {$k_T\leq 0$};
        \node at (0.9, -0.55) {$k_T\geq 0$};

        \path[pypurple, draw, line width = 0.8pt, postaction = decorate, 
		decoration={markings,
			mark=at position 0.5 with {\arrow[line width=1pt]{>}}}] (-1.5, 0) -- (-0.15,0);
            \path[pypurple, draw, line width = 0.8pt, postaction = decorate, 
		decoration={markings,
			mark=at position 0.5 with {\arrow[line width=1pt]{>}}}] (0.15,0) -- (1.5,0);
	\path[pyred, draw, dashed, line width = 0.8pt, postaction = decorate, 
		decoration={markings,
			mark=at position 0.25 with {\arrow[line width=1pt]{>}}}] (1.5,0.) arc (-3.81407:180+3.81407:1.50333);

	\path[pyblue, draw, dashed, line width = 0.8pt, postaction = decorate, 
		decoration={markings,
			mark=at position 0.25 with {\arrow[line width=1pt]{>}}}] (1.5,0) arc (-3.81407:-180+3.81407:1.50333);
	\path[pypurple, draw,  line width = 0.8pt, postaction = decorate, 
		decoration={markings,
			mark=at position 0.25 with {\arrow[line width=1pt]{>}}}] (-0.15,0) arc (180:0:0.15);
	\end{tikzpicture}
\end{center}
    \caption{The time integration contour for a dS scattering amplitude in an eternal energy-annihilating universe. The contour is chosen such that the integral on the large arc vanishes for $k_T\neq 0$ while the scattering amplitude is non-vanishing only when $k_T = k_{\text{in}} - k_{\text{out}}\geq 0$.}
    \label{EnergyCreation}
\end{figure}
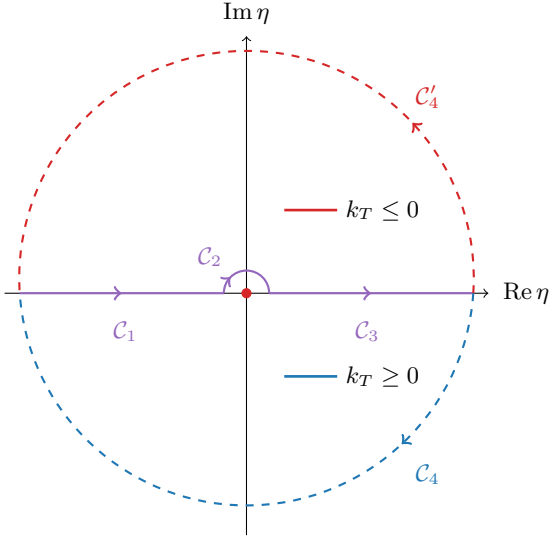

\section{Exceptional EFTs from generalised energy conservation}
We focus on imposing GEC on the four-point amplitudes arising from the self-interactions of exceptional series scalars to avoid potential instabilities. We could in principle impose a $\mathbb Z_2$ symmetry such that only contact diagrams contribute, however in each case we find that cubic vertices are forbidden by GEC and therefore no exchange contributions to four-point amplitudes are allowed. We will show this explicitly only for the $\Delta = 3$ scalar with the generalisation to other conformal dimensions simple. For four-point interactions we include operators with at most $2 \Delta-4$ derivatives. A benefit of working with this amplitude over e.g. boundary correlators is that it is field-redefinition invariant. This follows from its derivation via an LSZ reduction as we show explicitly in the Supplementary Material where we also work through two field-redefinition examples.

\paragraph{$\bullet$ $\Delta = 3$}
We begin our analysis with the first allowed mass of exceptional series scalars which is $m^2 = 0$ corresponding to $\Delta = 3$. The mode function is given by
 \begin{align} \label{MasslessModeFunction}
    f_k(\eta) = \frac{H e^{-ik\eta}}{\sqrt{2k^3}}(1 + ik\eta)\,.
\end{align}
Let's first consider cubic interactions where the only dS-invariant interaction that cannot be removed by a field redefinition is the potential $V(\phi) = c^{(3)}_{0} \phi^3$. The corresponding three-point amplitude is computed by 
\begin{align}
\mathcal{A}_3 = \frac{1}{2\pi}
   \frac{6 c^{(3)}_0}{H^4} \sqrt{8 k_1 k_2 k_3}
      \int_{-\infty}^{+\infty}\frac{d\eta}{\eta^4} f_{k_1}(\eta) f_{k_2}(\eta)f_{k_3}(\eta) \,.
\end{align}
In order to remove any potential instabilities, we demand that the residue of the integrand vanishes which leads to $\mathcal{A}_{k_T\neq 0}^{(\pm)} = 0$. Given that the Taylor expansion of $f_{k_1}(\eta) f_{k_2}(\eta)f_{k_3}(\eta)$ contains a non-zero $\eta^3$ part, $\mathcal{A}^{(\pm)}_{k_T\neq 0} = 0$ only if we set $c^{(3)}_0=0$. GEC therefore forbids any non-trivial cubic vertex for this massless scalar. 
 
Moving to quartic interactions, we consider the general Lagrangian up to two derivatives:
\begin{align}
     \frac{\mathcal{L}^{\Delta = 3}}{\sqrt{-g}} = -\frac{1}{2}(\partial\phi)^2 + d^{(3)}_{0} H^2 \phi^4  + d^{(3)}_2 \phi^2(\partial\phi)^2 +\ldots
\end{align}
We include a factor of $H^2$ in front of the potential such that the $d^{(\Delta)}_{2m}$ have the same non-zero mass dimension for each $\Delta$. The potential contributes to the residue, just like for the three-point amplitude, however the two-derivative term does not. This follows from the fact that the Taylor expansion of the mode function \eqref{MasslessModeFunction} does not contain a linear term in $\eta$ (which is the origin of the \textit{manifestly local test} of \cite{MLT}). We show this explicitly in the Supplementary Material where we discuss field redefinitions. To realise GEC we therefore need to fix $d^{(3)}_0 = 0$. 

We are therefore left with only the two-derivative interaction, but since this term can be removed by a field redefinition, as we discuss in the Supplementary Material, the amplitude in this case becomes trivial. To yield a non-zero result we can alter our starting point and allow for coloured massless scalars. In this case the potential $d^{(3)}_0 \phi^4$, which can now be coloured, is still forbidden by GEC but the two-derivative interaction is now non-trivial and does indeed yield non-vanishing amplitudes. In this sense, GEC picks out the four-point two-derivative interaction. If we wrote the interactions as $g^{IJ}(\phi) \partial^{\mu} \phi_{I} \partial_{\mu}\phi_{J}$, for some internal manifold metric $g^{IJ}(\phi)$, then non-trivial amplitudes arise when this symmetric metric has a non-vanishing Riemann tensor \cite{Cheung:2021yog}. This can be the case for two or more scalars.   

\paragraph{$\bullet$  $\Delta = 4$}We now turn our attention to a scalar with $m^2 = -4 H^2$ and $\Delta = 4$. Here the mode function is given by
\begin{align} \label{DBImodefunction}
    f_k(\eta) = \frac{H e^{-ik\eta}}{\sqrt{2k^5}\eta}[-3i + k\eta(3 + i k\eta)] \,.
\end{align}
In this case we immediately use our freedom to choose a field basis where there is no $\phi^2 (\partial \phi)^2$ operator (in the Supplementary Material we verify that this interaction can indeed be removed by a field redefinition) and now work with the most general quartic interactions up to four derivatives:
\begin{align} \label{FourDerivativeLagrangian}
    \frac{\mathcal{L}^{\Delta = 4}}{\sqrt{-g}} &= -\frac{1}{2} (\partial\phi)^2 + 2 H^2 \phi^2 + d^{(4)}_0 H^4 \phi^4 + d^{(4)}_4 (\partial \phi)^4 \ldots
\end{align}
We now compute the residue of the corresponding four-point amplitude and find 
\begin{align}
    &\mathcal{A}^{(\pm)}_4|_{k_T\neq 0} = -(d_0^{(4)} + 8 d_4^{(4)}) H^4 i\frac{36}{35}  \theta(\pm k_T) \nonumber\\
    &\frac{\biggl( 3 ( k_1^7 + k_2^7 +k_3^7 +k_4^7) - 7[k_1^2 k_2^2 (k_1^3 + k_2^3) + 5\text{perms} ] \biggr)}{k_1^2 k_2^2 k_3^2 k_4^2} \,,
\end{align}
where the sign in front of $k_T$ depends on the choice of contour shown in Figure \ref{EnergyAnnihilation} and Figure \ref{EnergyCreation}. We find that GEC requires us to fix $d^{(4)}_0 = -8 d^{(4)}_4$ which therefore fixes the interactions up to an overall coupling constant. It might seem surprising that operators with different numbers of derivatives can contribute to the amplitude in identical ways, but this is simply a consequence of scale invariance which yields a fixed momentum scaling. We know that there is indeed a special theory with $\Delta = 4$ and interactions with this power counting which is the DBI scalar (see e.g. \cite{Goon:2011uw, Bonifacio:2018zex, Bonifacio:2021mrf}) with leading-order Lagrangian:
\begin{align} \label{DBIaction}
\frac{\mathcal{L}^{\text{DBI}}}{\sqrt{-g}} = \frac{H^2 \Lambda^2}{(1 - \phi^2/\Lambda^2)^2}\sqrt{1 - \frac{(\partial \phi)^2 / (H^2 \Lambda^2)}{1 - \phi^2/\Lambda^2}} \,,
\end{align}
where $\Lambda$ is a dimensionful scale. Upon expanding this Lagrangian up to quartic order in $\phi$ and performing the field redefinition $\phi \rightarrow \phi - \phi^3/(2 \Lambda^2)$ we recover our action \eqref{FourDerivativeLagrangian} with $d^{(4)}_0 = -8 d^{(4)}_4$. GEC has therefore picked out the DBI scalar as being special amongst $\Delta = 4$ self-interacting scalars with at most one derivative per field. 

The full corresponding amplitude is quite long and not particularly illuminating but let us point out that the coefficient of the part of the amplitude with the highest number of derivatives acting $\delta(k_T)$ is precisely the flat-space amplitude, which is also the case for all $\Delta$. This is analogous to the statement that cosmological correlators computed at the boundary $\eta \rightarrow 0$ contain flat-space amplitudes as the coefficient of the leading total-energy pole \cite{Maldacena:2011nz,Raju:2012zr,BBBB,Pajer:2020wnj}.

\paragraph{$\bullet$ $\Delta = 5$} Next we consider a scalar with $m^2 = -10 H^2$ or equivalently $\Delta = 5$, where the mode function is given by 
\begin{align}
    f_\eta(k) = \frac{H e^{-ik\eta}}{\sqrt{2 k^7}\eta^2}\biggl\{ -15 - i k\eta\biggl[ 15 + k\eta (6i - k\eta) \biggr] \biggr\}.
\end{align}
In keeping with the previous pattern, in this case we allow for general interactions up to six derivatives, with the two-derivative operator removed by a field redefinition. There is only a single six-derivative operator that cannot be removed since the only Lorentz-invariant four-point amplitude in flat-space is $\mathcal{A} \sim st(s+t)$. We therefore work with
\begin{align}
    \frac{\mathcal{L}^{\Delta = 5}}{\sqrt{-g}} = &-\frac{1}{2}(\partial\phi)^2 + 5 H^2 \phi^2 + d^{(5)}_{0}H^6 \phi^4 \nonumber \\
    &+ d_{4}^{(5)} H^2 (\partial\phi)^4 + d_{6}^{(5)} (\partial\phi)^2 (\nabla_\mu\nabla_\nu\phi)^2 + \ldots
\end{align}
As before we now impose GEC by demanding that $\mathcal{A}_{k_{T} \neq0}^{(\pm)} = 0$ and find the conditions:
\begin{align}
d_4^{(5)} = -\frac{6}{125} d_0^{(5)},\;\; d_6^{(5)} = \frac{3}{500} d_0^{(5)}.
\end{align}
Note that in this case there are more independent kinematical structures compared to $\Delta=3,4$ which is why we find more conditions. 

Up to integration by parts and field redefinitions (which do not alter the amplitude), which are worked through in \cite{Armstrong:2022vgl}, these GEC conditions on the couplings uniquely pick out the Special Galileon (see \cite{Bonifacio:2018zex,Bonifacio:2021mrf} for details on this theory).
\paragraph{$\bullet$  $\Delta = 6$} We have so far used GEC to recover known theories of special scalars in dS space. We now initiate the search for new theories by increasing the conformal dimension. For $m^2 = -18 H^2$, or $\Delta = 6$, the mode function is given by
\begin{align}
    f_k(\eta) = &\frac{i H e^{-ik\eta}}{\sqrt{2 k^9} \eta^3}\biggl\{ 105 + k\eta\biggl[ 105i + \nonumber\\
    &k\eta\biggl( -45 + k\eta( -10i + k\eta) \biggr) \biggr] \biggr\}.
\end{align}
We now consider a general Lagrangian including operators up to eight derivatives. Again, there is a single new operator that cannot be removed by a field redefinition given that the only flat-space amplitude is $\mathcal{A} \sim (s^2 + t^2 + st)^2$ so with a choice of field basis we can write
\begin{align} \label{Delta6Lagrangian}
    \frac{\mathcal{L}^{\Delta = 6}}{\sqrt{-g}} =& -\frac{1}{2}(\partial\phi)^2 + 9 H^2 \phi^2 + d_{0}^{(6)} H^8 \phi^4 + d_{4}^{(6)} H^4 (\partial\phi)^4 \nonumber\\
    &+ d_{6}^{(6)} H^2(\partial\phi)^2 (\nabla_\mu\nabla_\nu\phi)^2 + d_{8}^{(6)} (\nabla_\mu\nabla_\nu\phi)^4  + \ldots
\end{align}
We now compute the four-point amplitude and by imposing GEC we find the conditions:
\begin{align}
    d_4^{(6)} = -\frac{13}{2646} d_0^{(6)},\;\; d_6^{(6)} = -\frac{5}{1323} d_0^{(6)},\;\; d_8^{(6)} = \frac{1}{5292}d_0^{(6)} \,. \label{Delta6Conditions}
\end{align}
To the best of our knowledge this theory has not been discussed before but given our results it is clearly a special subset of all possible $\Delta=6$ theories with at most $2 \Delta-4$ derivatives. Following the pattern for lower conformal dimensions, we might expect that this theory has a non-linearly realised symmetry. 

\paragraph{$\bullet$  $\Delta = 7$} We end our analysis by considering a scalar with $m^2 = -28 H^2$ or $\Delta = 7$, where the mode function is
\begin{align}
    f_k(\eta) = &\frac{H e^{-ik\eta}}{\sqrt{2k^{11}}\eta^4}\biggl\{ 945  + i k\eta\biggl[945 + k\eta\biggl(420i \nonumber\\
    &+ k\eta\biggl(-105 + k\eta(-15i + k\eta) \biggr) \biggr) \biggr] \biggr\}.
\end{align}
Continuing the previous pattern, we now allow for operators with at most ten derivatives. As before, there is a single new operator we can write down given that the only possible flat-space amplitude is $ \mathcal{A} \sim st(s+t)(s^2+t^2+st)$. A convenient basis for the Lagrangian is 
\begin{align} \label{Delta7Lagrangian}
\frac{\mathcal{L}^{\Delta = 7}}{\sqrt{-g}} =& -\frac{1}{2}(\partial\phi)^2 + 14 H^2 \phi^2 + d_0^{(7)}H^{10} \phi^4 + d_4^{(7)} H^6 (\partial\phi)^4 \nonumber\\
    &+ d_6^{(7)} H^4 (\partial\phi)^2 (\nabla_\mu\nabla_\nu\phi)^2 + d_8^{(7)} H^2 (\nabla_\mu\nabla_\nu\phi)^4 \nonumber\\
    &+ d_{10}^{(7)} \partial^{\alpha}[(\nabla_\mu\nabla_\nu \phi)^2] \partial_{\alpha}[(\nabla_\mu\nabla_\nu \phi)^2] + \ldots
\end{align}
 As with all our previous cases, the GEC condition uniquely fixes all couplings up to some overall coefficient:
\begin{align}
    &d_4^{(7)} = -\frac{129}{13034}d_0^{(7)},\;\;d_6^{(7)} = -\frac{141}{1981168}d_0^{(7)},\;\;\nonumber\\
    &d_8^{(7)} = \frac{3}{247646}d_0^{(7)},\;\; d_{10}^{(7)} = \frac{3}{7924672}d_0^{(7)}\,. \label{Delta7Conditions}
\end{align}
At this point, we conjecture that solutions to the GEC condition do not truncate as we increase the conformal dimension. We therefore expect to have infinitely many exceptional EFTs living in the exceptional series of de Sitter representations each with unique quartic self-interactions with at most $2 \Delta-4$ derivatives (but quite possibly requiring additional degrees of freedom for $\Delta \geq 6$).

\section{Discussion}

In this paper we have considered the dS S-matrix proposed in \cite{Donath:2024utn}, and by demanding the absence of any instabilities, we have constrained the self-interactions of exceptional series scalars. In such theories, the corresponding amplitudes only have support when energy is conserved. We rediscovered the DBI and Special Galileon theories, and found evidence for new exceptional theories each with uniquely fixed quartic self-interactions with at most $2 \Delta-4$ derivatives. The solutions we have found are therefore the leading-order ones in the derivative expansion. Our results lead us to make two conjectures: 
\begin{conjecture}
For each integer conformal dimension $\Delta \geq 4$, there is a unique four-point amplitude (allowing for operators with at most $2 \Delta-4$ derivatives) that only has support when energy is conserved.
\end{conjecture}
\begin{conjecture}
There is at least one exceptional EFT for each $\Delta\geq 4$ exceptional series representation of dS space. For $\Delta \geq 6$ such EFTs contain additional degrees of freedom. 
\end{conjecture}
Our second conjecture follows from the results of \cite{Bonifacio:2018zex} which suggests that \textit{single-scalar} exceptional EFTs do not exist for $\Delta \geq 6$. 

Interestingly, the exceptional EFTs are uniquely picked out by GEC/stability constraints in dS space, but not in AdS, since energy is automatically conserved in AdS by exact time-translation symmetry. In that sense, no tuning of the interactions is required to yield healthy theories in AdS. This conclusion coincides with representation theory, where in dS the exceptional EFTs live in the exceptional series of dS representations, whereas in AdS, they simply correspond to conventional scalar representations \cite{Bonifacio:2018zex}. We further point out that the uniqueness of the theories we have fixed using GEC, especially DBI and Special Galileon, which involve only a single scalar degree of freedom, resonates with the uniqueness of Yang-Mills and General Relativity in flat-space as theories of massless spin-$1$ and spin-$2$ particles, respectively. Indeed, there we start with a particular representation of the Poincar\'{e} group (massless spin-$1$ or massless spin-$2$) and demand the absence of instability. To some order in derivatives, the condition yields unique theories (see e.g. \cite{ Deser:1969wk, deRham:2014zqa}). Here, we start with particular representations of the dS group, and by demanding the absence of instabilities, we arrive at unique theories, again up to some order in derivatives. Therefore, the exceptional theories in dS space appear more fundamental than their flat space counterparts, since additional constraints such as soft limits \cite{Cheung:2014dqa,Cheung:2016drk,Cheung:2015ota,Padilla:2016mno}, or non-linear symmetries \cite{Roest:2019oiw,Bogers:2018zeg,Hinterbichler:2015pqa} are required to uniquely fix the latter, beyond representation theory and stability alone. In any case, it is interesting to study the connections between GEC and non-linearly realised symmetries in de Sitter space, and recently it was shown in \cite{Du:2025glv} that solutions to the soft theorems that follow from the non-linear symmetries require the imposition of GEC. 

We hope that our work initiates a systematic search for new theories in dS space and a more ``on-shell" study of special theories. We envisage a number of possible future research directions:

\begin{itemize}
\item Moving to six-point might be an avenue for realising that new degrees of freedom are necessary for $\Delta \geq 6$. Indeed, it could be that exchange and contact diagrams due to scalar self-interactions cannot combine into six-point amplitudes that adhere to GEC and new degrees of freedom, and therefore new exchange processes, are required. 
\item It would be interesting to systematically search for extensions of the dS algebra whose non-linear realisation contains the scalars we have discovered in this work. 
   \item The infinite towers of theories we have found could correspond to the longitudinal modes of the infinite higher spin fields in (A)dS in the decoupling limit \cite{Joung:2012rv, Alkalaev:2011zv, Brust:2016zns}. It would be interesting to investigate this, and other connections to partially-massless spinning fields, in more detail. In this regard the spinor helicity formalism of \cite{Basile:2024ydc} could be very useful.
    \item Demanding a vanishing residue for the integrands we have studied in this work is equivalent to demanding the absence of logarithmic divergences in boundary correlators. The amplitude has the benefit of being field-redefinition invariant, nevertheless it would be interesting to study the properties of boundary correlators for these exceptional theories in more detail \footnote{K. Pan, Z. Qin, and Z.-Z. Xianyu, {\it to appear.}}.
\end{itemize}

\section*{Acknowledgements}

We thank Trevor Cheung, Kurt Hinterbichler, Enrico Pajer, Sadra Jazayeri, Kezhao Pan, Diederik Roest, Xi Tong, Kieran Wood, Zhong-Zhi Xianyu and Yuhang Zhu for useful discussions. Special thanks to Qixin-Xie for helping us with some of the numerical implementation. Both authors would also like to express their sincere appreciation to Zhe-Han Qin for insightful discussions and his early involvement in the project. Z.D. is supported by Nottingham CSC [file No. 202206340007].
D.S. is supported by a UKRI Stephen Hawking
Fellowship [grant number EP/W005441/1] and a Nottingham Research Fellowship from the University
of Nottingham.

For the purpose of open access, the authors have applied a CC BY public copyright licence to any Author
Accepted Manuscript version arising.

\paragraph{Data access statement} No new data were
created or analysed during this study. 

\appendix

\section{A de Sitter S-matrix and field basis independence}
In this Appendix we briefly discuss the dS S-matrix proposed in \cite{Donath:2024utn} and show that it follows from an LSZ reduction. 

To realise asymptotic BD vacua in both the past and future, the authors of \cite{Donath:2024utn} suggest to glue together two Poincar\'{e} patches of dS space. The metric is then given by
\begin{align}
    ds^2 = \frac{-d\eta^2 + d\vec{x}^2}{H^2 \eta^2}, \;\;\eta\in(-\infty,+\infty)\,. \label{Metric}
\end{align}
In-out correlation functions on this spacetime are defined as
\begin{align}
    G(\eta, \vec{x}) = {}_{\text{out}} {\braket{\Omega|\mathcal{T}\{\phi(x)\phi(x_1)\ldots\phi(x_n)\}|\Omega}}_{\text{in}} \,,
\end{align}
and for IR-finite interactions these in-out correlators are equivalent to the familiar in-in correlators of cosmology \cite{Donath:2024utn}.
Here ${}_{\text{out}}{\bra{\Omega}}$ and $\ket{\Omega}_{\text{in}}$ are defined by evolving the free BD vacua $\bra{0}$ and $\ket{0}$ from the far future and far past, respectively, and $\mathcal{T}$ is the time-ordering operator. We keep the dependence on $x_1$ to $x_n$ implicit for simplicity. A corresponding S-matrix can be obtained by amputating the external legs using the LSZ reduction in the same way as in flat space. To see how the LSZ operator arises, we perform a Fourier transform of coordinates $x$ on the in-out correlator analogous to \cite{Melville:2023kgd}:
\begin{align}
    \tilde{G}(k^0,\vec{k}) \equiv \int\frac{d\eta}{H\eta} d^3x e^{i(k^0\eta - \vec{k}\cdot\vec{x})} G(\eta, \vec{x}) \,.
\end{align}
Note that we include a factor of $1/\eta$ in the Fourier transform to account for the fact we are not working with a canonically normalised field c.f. \eqref{MasslessModeFunc}. We split the integral into three separate regions
\begin{align}
    &(a)\;\;\;\eta>T^+,\\
    &(b)\;\;\;T^+\geq\eta\geq T^-,\\
    &(c)\;\;\;T^->\eta \,,
\end{align}
where $T^+$ and $T^-$ are defined such that for region $(a)$ (in the far future) and $(c)$ (in the far past), wavepackets are well separated. The mode expansion of the field $\phi(x)$ is given by
\begin{align}
        \phi(x) = \int \frac{d^3 k}{(2\pi)^3} f_k(\eta) e^{i\vec{k}\cdot\vec{x}} a_{\vec{k}} + c.c.,
\end{align}
where $a_{\vec{k}}$ is the annihilation operator, and the BD mode function is given by
\begin{align}
    f_k(\eta) = \frac{\sqrt{\pi}H}{2} (-\eta)^{\frac{3}{2}} e^{i\frac{\pi}{2}(\nu - 1/2)} H_{\nu}^{(1)}(-k\eta),
\end{align}
where $\nu = \Delta - \frac{3}{2}$ and $H_{\nu}^{(1)}(x)$ is the Hankel function of the first kind. Note that for our interests with integer $\Delta \geq 3$, the mode function simplifies and has no branch cut. 

Let's first consider the integral in the far-future region where any exceptional series scalar field (with BD normalisation) can be approximated by
\begin{align}
    \phi(x) \stackrel{\eta>T^+}{\approx} \int \frac{d^3 k}{(2\pi)^3\sqrt{2k}} (i H \eta)e^{-ik\eta +i\vec{k}\cdot\vec{x}}a_{\vec{k}} + c.c. \label{MasslessModeFunc}\,.
\end{align}
In the far future, the creation operator annihilates the out vacuum, while the annihilation operator creates a one-particle state $\langle\vec k|$ in the out state. The integral therefore reads
\begin{align}
    \tilde{G}^+(k^0,\vec{k}) &\equiv \int_{\eta>T^+}\frac{d\eta}{H\eta} d^3x  e^{i(k^0\eta - \vec{k}\cdot\vec{x})} G(\eta,\vec{x}) \nonumber\\
    &= -\frac{ e^{i(k^0-k)T^+}}{2k(k^0-k)}  \braket{\vec{k}|\mathcal{T}\{\phi(x_1)\ldots\phi(x_n)\}|\Omega}_{in} \,,
\end{align}
which has a similar pole structure to the flat-space correlation function. Therefore, the dS S-matrix for those scalars can be obtained by amputating all the external legs:
\begin{align}
    S_{\beta\alpha}&\nonumber\\
    = (-1)^n&\prod_{a\in \text{out}}\lim_{k_a^0\rightarrow k_a} ({k_a^0}^2-k_a^2)\int\frac{d\eta_a}{H\eta_a} d^3x e^{i(k^0_a\eta_a - \vec{k}_a\cdot\vec{x}_a)}\nonumber\\
    &\prod_{b \in \text{in}}\lim_{k_b^0\rightarrow k_b} ({k_b^0}^2-k_b^2)\int\frac{d\eta_b}{H\eta_b} d^3x e^{-i(k^0_b\eta_b - \vec{k}_b\cdot\vec{x}_b)}\nonumber\\
    &{}_{\text{out}} {\braket{\Omega|\mathcal{T}\{\phi(x_1)\ldots\phi(x_n)\}|\Omega}}_{\text{in}}.
\end{align}
The dS S-matrix is computed by the Dyson series as explained in the main text. Given the LSZ reduction formula, the S-matrix is field-redefinition invariant and unaffected by boundary terms, as in flat space. It is constructive, however, to see via explicit examples that field redefinitions do not change the S-matrix. We will show an example for both $\Delta=3$ and $\Delta=4$.

\paragraph{$\bullet$  $\Delta = 3$} Consider a massless scalar where the free theory after the field redefinition $\phi\rightarrow \phi - \frac{\lambda}{3}\phi^3$ becomes
\begin{align} \label{FieldRedefAction}
    S=\int \sqrt{-g}\biggl[-\frac{1}{2}(\partial\phi)^2 + \lambda\phi^2(\partial\phi)^2 \biggr] \,,
\end{align}
where $\lambda$ is a dimensionful constant. 
The tree-level four-point S-matrix is (the reader should keep in mind that when we compute a $4 \rightarrow 0$ amplitude we are really computing a $2 \rightarrow 2$ amplitude after crossing) 
\begin{align}
    &\braket{0|S|\vec{k}_1, \vec{k}_2, \vec{k}_3, \vec{k}_4} =
    \sqrt{16 k_1 k_2 k_3 k_4}\delta^3 \left(\sum_{a=1}^4 \vec{k}_a \right)\nonumber\\
    & \frac{-\lambda}{H^2} \int_{-\infty}^{+\infty}\frac{d\eta}{\eta^2}\biggl[\partial_\eta f_{k_1}(\eta)\partial_{\eta}f_{k_2}(\eta) \nonumber\\
    &+ (\vec{k}_1\cdot\vec{k}_2) f_{k_1}(\eta)f_{k_2}(\eta)\biggr]f_{k_3}(\eta)f_{k_4}(\eta) + 23\;\text{perms} \,,
\end{align}
where the mode function is given by 
 \begin{align} \label{MasslessModeFunction}
    f_k(\eta) = \frac{H e^{-ik\eta}}{\sqrt{2k^3}}(1 + ik\eta)\,.
\end{align}
The interactions are adiabatically turned on/off via shifting the Hamiltonian by $\mathcal{H}_{\text{int}}\rightarrow e^{-\epsilon |\eta|}\mathcal{H}_{\text{int}}$ \cite{Donath:2024utn, Gell-Mann:1951ooy}. Such a shift ensures convergence of the time integral at both past and future infinity. To deal with the pole at $\eta=0$, the integral along the real axis is deformed. We then divide the integrand into two pieces based on the overall power of $\eta$:
\begin{align}
\text{Contribution 1}: ~~ &\int^{+ \infty}_{- \infty} d \eta  \eta^{p} e^{- i k_{T} \eta}, \qquad p < 0 \,, \\
\text{Contribution 2}: ~~ &\int^{+ \infty}_{- \infty} d \eta \eta^{p} e^{- i k_{T} \eta}, \qquad p \geq 0  \,.
\end{align}

First consider contribution 1 with $k_{T} \neq 0$. For $k_{T} < 0$ (for our current example we have $k_T \geq 0$ but more generally $k_T \leq 0$ is possible) we close the contour by a large arc in the upper-half of the complex $\eta$-plane, while for $k_T>0$ we close the contour in the lower-half complex plane. The two possible $i \epsilon$ deformations, along with the closure of the contours, are shown in Figures 1 and 2 of the main text. For definiteness, let us work with $k_{T} > 0$. For the contour in Figure 2 of the main text, where the contour encloses the pole, the integral along the closed contour is fixed by the residue at $\eta=0$, while for the contour in Figure 1 of the main text where the pole is not enclosed by the contour, the integral vanishes. In both cases, the contribution from the large arc at infinity vanishes due to our BD boundary conditions and therefore the integral along the real line can be computed by the residue theorem. It is straightforward to compute the residue at $\eta=0$ by performing a Laurent expansion of the integrand.
 The coefficient of the $1 / \eta$ term in the expansion has a vanishing coefficient and therefore the residue vanishes for each permutation separately. We note that the vanishing of the residue arises for the same reason massless correlators satisfy the \textit{manifest locality test} (MLT) of \cite{MLT}. For both $i \epsilon$ prescriptions we therefore find a vanishing answer for contribution 1 when $k_T \neq 0$.

For $k_{T} = 0$ we need to verify that the arc at infinity still yields a vanishing result. Given that the residue vanishes, we know that the coefficient of the $p = -1$ term is proportional to $k_{T}$ and therefore the large arc again vanishes. We have therefore shown that contribution 1 vanishes for both contour choices and therefore we have $\mathcal{A}^{(\pm)}_{k_{T} \neq 0} = 0$ by the residue theorem.

Contribution 2 is straightforward to compute given the absence of any possible poles and we get contributions for $p = 0,1,2$. We find
\begin{align} \label{FieldRedefAmp}
\mathcal{A} = \lambda H^2 [  \mathcal{A}^{(0)} \delta(k_{T}) + \mathcal{A}^{(1)} \partial_{k_{T}} \delta(k_{T}) + \mathcal{A}^{(2)}\partial^{2}_{k_{T}} \delta(k_{T})]\,,
\end{align}
where 
\begin{align}
    \mathcal{A}^{(0)} &= \frac{-2}{k_1 k_2 k_3 k_4}[(k_2 + k_3 + k_4)(k_1^3 + k_1(k_2^2 + k_3^2 + k_4^2) \nonumber\\
    &+ (k_2^2 + k_3 k_4)(k_3 + k_4) + k_2 ( k_3^2 - k_3 k_4 + k_4^2 )) \nonumber\\
    &+ k_1^2 ( 2 k_2^2 + 2 k_3^2 + 2 k_4^2 + k_2 k_3 + k_2 k_4 + k_3 k_4) ], 
\end{align}
\begin{align}
    \mathcal{A}^{(1)}  &= \frac{-2}{ k_1 k_2 k_3 k_4} (k_1+k_2+k_3+k_4) [k_2 k_3 k_4 (k_2 + k_3 + k_4) \nonumber\\
    &+ k_1^2( k_2 k_3 + k_2 k_4 + k_3 k_4 ) +  k_1 k_2 (k_3 - k_4)^2 \nonumber \\ &+ k_1 k_2^2( k_3 + k_4) + k_1 k_3 k_4 (k_3+k_4)], 
\end{align}
\begin{align}
    \mathcal{A}^{(2)}  &= -2(k_1+k_2+k_3+k_4)^2 \,.
\end{align}
To arrive that these expressions we have summed over all permutations and have eliminated $\vec{k}_{a} \cdot \vec{k}_{b}$ products with $a \neq b$ by momentum conservation. To properly analyse \eqref{FieldRedefAmp} we need to factorise the delta functions. We can write 
\begin{align} 
\frac{\mathcal{A}}{\lambda H^2} 
& = (\mathcal{A}^{(0)}|_{k_{T} = 0} - \partial_{k_{T}}\mathcal{A}^{(1)} |_{k_{T} = 0} + \partial^2_{k_{T}}\mathcal{A}^{(2)} |_{k_{T} = 0})\delta(k_{T}) \nonumber \\
& + (\mathcal{A}^{(1)}|_{k_{T} = 0} - 2 \partial_{k_{T}}\mathcal{A}^{(2)}|_{k_{T} = 0})\partial_{k_{T}} \delta(k_{T}) \nonumber \\
& + \mathcal{A}^{(2)}|_{k_{T} = 0}\partial^{2}_{k_{T}} \delta(k_{T}) \,. \label{FieldRedefAmpExpanded}
\end{align}
From the above explicit expressions we have 
\begin{align}
\mathcal{A}^{(0)}|_{k_{T} = 0}  &= -8, \qquad 
\partial_{k_{T}}\mathcal{A}^{(1)} |_{k_{T} = 0} = -12 \label{AmplitudeCoefficients1} \,,\\
\partial^2_{k_{T}}\mathcal{A}^{(2)} |_{k_{T} = 0} &= -4, ~~~~~ \qquad
\mathcal{A}^{(1)}|_{k_{T} = 0}  = 0 \,, \label{AmplitudeCoefficients2} \\
\partial_{k_{T}}\mathcal{A}^{(2)} |_{k_{T} = 0} &= 0, \qquad
~~~~~~~\mathcal{A}^{(2)} |_{k_{T} = 0} = 0 \,, \label{AmplitudeCoefficients3}
\end{align}
and therefore it is simple to see that the four-point amplitude arising from \eqref{FieldRedefAction} vanishes. We note that just like in flat space, we need to sum over all permutations to see that the amplitude vanishes.

After partial integration we can bring the vertex $\phi^2 (\partial \phi)^2$ into the form $\phi^3 \Box \phi$ which manifestly vanishes on-shell and therefore yields a vanishing amplitude. However, this requires us to drop a boundary term. In flat-space QFT we are used to dropping such boundary terms, but the situation is more subtle in cosmology since we know that boundary terms can indeed alter the form of boundary correlators. What we are therefore showing here, is that the boundary term does not contribute to the dS amplitude. This is to be expected given the LSZ reduction, however we find it illuminating to see how things work out in practice. 

\paragraph{$\bullet$  $\Delta = 4$} We further check such invariance at the level of four-point contact diagrams for a scalar with mass $m^2 = -4 H^2$ as relevant for the DBI scalar. If we take the free theory and perform the field redefinition $\phi \rightarrow \phi - \frac{\lambda}{3}\phi^3$ then the action is 
\begin{align} \label{DBIFieldRedefAction}
    S=\int \sqrt{-g}\biggl[-\frac{1}{2}(\partial\phi)^2 + 2 H^2 \phi^2 -\frac{4}{3} \lambda H^2 \phi^4 + \lambda \phi^2 (\partial \phi)^2\biggr].
\end{align}
The four-point S-matrix is
\begin{align}
    &\braket{0|S|\vec{k}_1, \vec{k}_2, \vec{k}_3, \vec{k}_4} =
    \sqrt{16 k_1 k_2 k_3 k_4}\delta^3 \left(\sum_{a=1}^4 \vec{k}_a \right)\nonumber\\
    & \frac{-\lambda}{H^2} \int_{-\infty}^{+\infty}\frac{d\eta}{\eta^2}\biggl[\partial_\eta f_{k_1}(\eta)\partial_{\eta}f_{k_2}(\eta) + (\vec{k}_1\cdot\vec{k}_2) f_{k_1}(\eta)f_{k_2}(\eta)  \nonumber\\
    & - \frac{4}{3 \eta^2} f_{k_1}(\eta)f_{k_2}(\eta)\biggr]f_{k_3}(\eta)f_{k_4}(\eta) + 23\;\text{perms} \,,
\end{align}
where the mode function is now given by
\begin{align} \label{DBImodefunction}
    f_k(\eta) = \frac{H e^{-ik\eta}}{\sqrt{2k^5}\eta}[-3i + k\eta(3 + i k\eta)] \,.
\end{align}
We start by computing the residue at $\eta=0$ and find that it vanishes which relies on \textit{both} the tuning between the two quartic interactions in \eqref{DBIFieldRedefAction} and summing over all permutations. For any other tuning between these interactions, the residue is non-zero. To complete the computation of contribution 1 for this amplitude we need to consider the contributions at $k_T = 0$ and again we know that the $p=-1$ coefficient is proportional to $k_{T}$ so the large arc still yields a vanishing result. Contribution 1 is therefore zero.

Moving to contribution 2, we again have contributions from $p=0,1,2$ and can therefore write the amplitude as \eqref{FieldRedefAmpExpanded}. Even though the full amplitudes for $\Delta=3$ and $\Delta=4$ are very different, we find that for $\Delta=4$ they still obey \eqref{AmplitudeCoefficients1}, \eqref{AmplitudeCoefficients2} and \eqref{AmplitudeCoefficients3} and therefore the full amplitude again vanishes.  

For the massless case the vanishing of the amplitude required a sum over permutations, while for the $\Delta=4$ scalar the vanishing required a sum over permutations and a collaboration between two different interactions that are simultaneously produced by the field redefinition. 

\section{Application to other dimensions}

In the main body we focused on field theories in four spacetime dimensions. Here we will show that GEC is equally powerful in other dimensions too. We will not provide a complete analysis, rather our aim is to offer a proof of principle that GEC is applicable to other dimensions.

In general spacetime dimension $D$, the mode function of a massive scalar field with Bunch-Davies vacuum conditions is
\begin{align}
   f_k(\eta) = \frac{\sqrt{\pi}H^{\frac{D-2}{2}}}{2} (-\eta)^{\frac{D-1}{2}} e^{i\frac{\pi}{2}(\nu - 1/2)} H_{\nu}^{(1)}(-k\eta),
\end{align}
where $\nu = \Delta - \frac{D-1}{2}$, and the conformal dimension is related to the mass by
\begin{align}
    \Delta = \frac{D-1}{2}+\sqrt{\frac{(D-1)^2}{4}-\frac{m^2}{H^2}} \,.
\end{align}
We now demonstrate the applicability of GEC for $D \neq 4$ by considering a scalar with $\Delta= D$ and $m^2 = -D H^2$, showing that the GEC condition precisely reproduces the DBI interactions for $D=2$ and $D=6$. We work with even spacetime dimensions since there we have discrete series scalars with mode functions that do not have a branch cut. We work with the same Lagrangian as in the main text i.e.
\begin{align} \label{FourDerivativeLagrangian}
    \frac{\mathcal{L}^{\Delta = D}}{\sqrt{-g}} &= -\frac{1}{2} (\partial\phi)^2 + \frac{D}{2} H^2 \phi^2 + d^{(4)}_0 H^4 \phi^4 + d^{(4)}_4 (\partial \phi)^4 \ldots
\end{align}
where the dimension of the coefficients are given by $[d^{(4)}] = -D$.
\paragraph{$\bullet$ $D = 2$}
The mode function for a $\Delta = 2$ scalar in $D=2$ is given by
\begin{align}
        f_k(\eta) =- \frac{ e^{-ik\eta}}{\sqrt{2k^3}\eta}(1 + ik\eta)\,.
\end{align}
The non-energy conserving part of the four-point amplitude is given by
\begin{align}
&\mathcal{A}^{(\pm)}_4|_{k_T\neq 0} = -(3d_0^{(4)} + 8 d_4^{(4)}) H^2 i\frac{4}{15}  \theta(\pm k_T) \nonumber\\
    &\frac{\biggl(  ( k_1^5 + k_2^5 +k_3^5 +k_4^5) - 5[k_1^2 k_2^2 (k_1 + k_2) + 5\text{perms} ] \biggr)}{k_1 k_2 k_3 k_4} \,.
\end{align}
The GEC condition therefore imposes $d_4^{(4)} = -\frac{3}{8} d_0^{(4)}$, which precisely matches the structure of DBI action in $D=2$ i.e.\cite{Bonifacio:2021mrf}
\begin{align} \label{DBIaction}
\frac{\mathcal{L}^{\text{DBI}}}{\sqrt{-g}} = \frac{H^2 }{(1 - \phi^2)}\sqrt{1 - \frac{(\partial \phi)^2 /H^2}{1 - \phi^2}} \,,
\end{align}
upon performing the field redefinition $\phi\rightarrow \phi - \phi^3/3$. Notice the theory introduces no extra scale beyond the Hubble scale since in $D=2$ the canonically normalised scalar is dimensionless.

\paragraph{$\bullet$ $D = 6$}
The mode function for a $\Delta = 6$ scalar in $D=6$ is given by
\begin{align}
         f_\eta(k) = \frac{H^2 e^{-ik\eta}}{\sqrt{2 k^7}\eta}\biggl\{ -15 - i k\eta\biggl[ 15 + k\eta (6i - k\eta) \biggr] \biggr\}.
\end{align}
The non-energy conserving part of the four-point amplitude is then given by
\begin{align}
&\mathcal{A}^{(\pm)}_4|_{k_T\neq 0} = -(d_0^{(4)} + 16 d_4^{(4)}) H^6 i\frac{60}{7}  \theta(\pm k_T) \nonumber\\
    &\frac{\biggl(  5( k_1^9 + k_2^9 +k_3^9 +k_4^9) - 9[k_1^2 k_2^2 (k_1^5 + k_2^5) + 5\text{perms} ] \biggr)}{k_1^3 k_2^3 k_3^3 k_4^3} \,.
\end{align}
The GEC condition therefore imposes $d_4^{(4)} = -\frac{1}{16} d_0^{(4)}$. The DBI action in $D=6$ takes the following form \cite{Bonifacio:2021mrf}
\begin{align} \label{DBIactionD=6}
\frac{\mathcal{L}^{\text{DBI}}}{\sqrt{-g}} = \frac{H^2 \Lambda^4}{(1 - \phi^2/\Lambda^4)^3}\sqrt{1 - \frac{(\partial \phi)^2 / (H^2 \Lambda^4)}{1 - \phi^2/\Lambda^4}} \,,
\end{align}
and upon performing the field redefinition $\phi\rightarrow \phi - \frac{2}{3\Lambda^4}\phi^3$ we see that the structure is consistent with tunings between operators forced by GEC. 

We have therefore shown that GEC can be used to pick out special theories in other spacetime dimensions thereby further illustrating its power at the level of the de Sitter S-matrix we have studied in this work.

\bibliographystyle{apsrev4-1}
\bibliography{RefDBI}

@article{deRham:2014zqa,
    author = "de Rham, Claudia",
    title = "{Massive Gravity}",
    eprint = "1401.4173",
    archivePrefix = "arXiv",
    primaryClass = "hep-th",
    doi = "10.12942/lrr-2014-7",
    journal = "Living Rev. Rel.",
    volume = "17",
    pages = "7",
    year = "2014"
}

@article{Deser:1969wk,
    author = "Deser, Stanley",
    title = "{Selfinteraction and gauge invariance}",
    eprint = "gr-qc/0411023",
    archivePrefix = "arXiv",
    doi = "10.1007/BF00759198",
    journal = "Gen. Rel. Grav.",
    volume = "1",
    pages = "9--18",
    year = "1970"
}

@article{Gell-Mann:1951ooy,
    author = "Gell-Mann, Murray and Low, Francis",
    title = "{Bound states in quantum field theory}",
    doi = "10.1103/PhysRev.84.350",
    journal = "Phys. Rev.",
    volume = "84",
    pages = "350--354",
    year = "1951"
}

@article{Brust:2016zns,
    author = "Brust, Christopher and Hinterbichler, Kurt",
    title = "{Partially Massless Higher-Spin Theory}",
    eprint = "1610.08510",
    archivePrefix = "arXiv",
    primaryClass = "hep-th",
    doi = "10.1007/JHEP02(2017)086",
    journal = "JHEP",
    volume = "02",
    pages = "086",
    year = "2017"
}

@article{Noether:1918zz,
    author = "Noether, Emmy",
    title = "{Invariant Variation Problems}",
    eprint = "physics/0503066",
    archivePrefix = "arXiv",
    doi = "10.1080/00411457108231446",
    journal = "Gott. Nachr.",
    volume = "1918",
    pages = "235--257",
    year = "1918"
}

@article{Melville:2023kgd,
    author = "Melville, Scott and Pimentel, Guilherme L.",
    title = "{de Sitter S matrix for the masses}",
    eprint = "2309.07092",
    archivePrefix = "arXiv",
    primaryClass = "hep-th",
    doi = "10.1103/PhysRevD.110.103530",
    journal = "Phys. Rev. D",
    volume = "110",
    number = "10",
    pages = "103530",
    year = "2024"
}

@article{Burrage:2011bt,
    author = "Burrage, Clare and de Rham, Claudia and Heisenberg, Lavinia",
    title = "{de Sitter Galileon}",
    eprint = "1104.0155",
    archivePrefix = "arXiv",
    primaryClass = "hep-th",
    doi = "10.1088/1475-7516/2011/05/025",
    journal = "JCAP",
    volume = "05",
    pages = "025",
    year = "2011"
}

@article{Alkalaev:2011zv,
    author = "Alkalaev, Konstantin and Grigoriev, Maxim",
    title = "{Unified BRST approach to (partially) massless and massive AdS fields of arbitrary symmetry type}",
    eprint = "1105.6111",
    archivePrefix = "arXiv",
    primaryClass = "hep-th",
    reportNumber = "FIAN-TD-2011-08",
    doi = "10.1016/j.nuclphysb.2011.08.005",
    journal = "Nucl. Phys. B",
    volume = "853",
    pages = "663--687",
    year = "2011"
}

@article{Joung:2012rv,
    author = "Joung, Euihun and Lopez, Luca and Taronna, Massimo",
    title = "{On the cubic interactions of massive and partially-massless higher spins in (A)dS}",
    eprint = "1203.6578",
    archivePrefix = "arXiv",
    primaryClass = "hep-th",
    doi = "10.1007/JHEP07(2012)041",
    journal = "JHEP",
    volume = "07",
    pages = "041",
    year = "2012"
}

@article{Donath:2024utn,
    author = "Donath, Yaniv and Pajer, Enrico",
    title = "{The in-out formalism for in-in correlators}",
    eprint = "2402.05999",
    archivePrefix = "arXiv",
    primaryClass = "hep-th",
    doi = "10.1007/JHEP07(2024)064",
    journal = "JHEP",
    volume = "07",
    pages = "064",
    year = "2024"
}

@article{Garoffolo:2025igz,
    author = "Garoffolo, Alice and Hinterbichler, Kurt and Trodden, Mark",
    title = "{Multi-Galileons in Curved Space}",
    eprint = "2505.08865",
    archivePrefix = "arXiv",
    primaryClass = "hep-th",
    month = "5",
    year = "2025"
}

@article{Cheung:2021yog,
    author = "Cheung, Clifford and Helset, Andreas and Parra-Martinez, Julio",
    title = "{Geometric soft theorems}",
    eprint = "2111.03045",
    archivePrefix = "arXiv",
    primaryClass = "hep-th",
    reportNumber = "CALT-TH-2021-038",
    doi = "10.1007/JHEP04(2022)011",
    journal = "JHEP",
    volume = "04",
    pages = "011",
    year = "2022"
}

@article{Roest:2019oiw,
    author = "Roest, Diederik and Stefanyszyn, David and Werkman, Pelle",
    title = "{An Algebraic Classification of Exceptional EFTs}",
    eprint = "1903.08222",
    archivePrefix = "arXiv",
    primaryClass = "hep-th",
    doi = "10.1007/JHEP08(2019)081",
    journal = "JHEP",
    volume = "08",
    pages = "081",
    year = "2019"
}

@article{Pajer:2020wnj,
    author = "Pajer, Enrico and Stefanyszyn, David and Supe\l{}, Jakub",
    title = "{The Boostless Bootstrap: Amplitudes without Lorentz boosts}",
    eprint = "2007.00027",
    archivePrefix = "arXiv",
    primaryClass = "hep-th",
    doi = "10.1007/JHEP12(2020)198",
    journal = "JHEP",
    volume = "12",
    pages = "198",
    year = "2020",
    note = "[Erratum: JHEP 04, 023 (2022)]"
}

@article{Cheung:2016drk,
    author = "Cheung, Clifford and Kampf, Karol and Novotny, Jiri and Shen, Chia-Hsien and Trnka, Jaroslav",
    title = "{A Periodic Table of Effective Field Theories}",
    eprint = "1611.03137",
    archivePrefix = "arXiv",
    primaryClass = "hep-th",
    reportNumber = "CALT-TH-2016-032",
    doi = "10.1007/JHEP02(2017)020",
    journal = "JHEP",
    volume = "02",
    pages = "020",
    year = "2017"
}

@article{Armstrong:2022vgl,
    author = "Armstrong, Connor and Lipstein, Arthur and Mei, Jiajie",
    title = "{Enhanced soft limits in de Sitter space}",
    eprint = "2210.02285",
    archivePrefix = "arXiv",
    primaryClass = "hep-th",
    doi = "10.1007/JHEP12(2022)064",
    journal = "JHEP",
    volume = "12",
    pages = "064",
    year = "2022"
}

@article{Hinterbichler:2015pqa,
    author = "Hinterbichler, Kurt and Joyce, Austin",
    title = "{Hidden symmetry of the Galileon}",
    eprint = "1501.07600",
    archivePrefix = "arXiv",
    primaryClass = "hep-th",
    doi = "10.1103/PhysRevD.92.023503",
    journal = "Phys. Rev. D",
    volume = "92",
    number = "2",
    pages = "023503",
    year = "2015"
}

@article{Penedones:2023uqc,
    author = "Penedones, Joao and Salehi Vaziri, Kamran and Sun, Zimo",
    title = "{Hilbert space of Quantum Field Theory in de Sitter spacetime}",
    eprint = "2301.04146",
    archivePrefix = "arXiv",
    primaryClass = "hep-th",
    month = "1",
    year = "2023"
}

@article{Bittermann:2022nfh,
    author = "Bittermann, Noah and Joyce, Austin",
    title = "{Soft limits of the wavefunction in exceptional scalar theories}",
    eprint = "2203.05576",
    archivePrefix = "arXiv",
    primaryClass = "hep-th",
    month = "3",
    year = "2022"
}

@article{MLT,
    author = "Jazayeri, Sadra and Pajer, Enrico and Stefanyszyn, David",
    title = "{From Locality and Unitarity to Cosmological Correlators}",
    eprint = "2103.08649",
    archivePrefix = "arXiv",
    primaryClass = "hep-th",
    month = "3",
    year = "2021"
}

@article{Padilla:2016mno,
    author = "Padilla, Antonio and Stefanyszyn, David and Wilson, Toby",
    title = "{Probing Scalar Effective Field Theories with the Soft Limits of Scattering Amplitudes}",
    eprint = "1612.04283",
    archivePrefix = "arXiv",
    primaryClass = "hep-th",
    doi = "10.1007/JHEP04(2017)015",
    journal = "JHEP",
    volume = "04",
    pages = "015",
    year = "2017"
}

@article{Cheung:2014dqa,
    author = "Cheung, Clifford and Kampf, Karol and Novotny, Jiri and Trnka, Jaroslav",
    title = "{Effective Field Theories from Soft Limits of Scattering Amplitudes}",
    eprint = "1412.4095",
    archivePrefix = "arXiv",
    primaryClass = "hep-th",
    reportNumber = "CALT-TH-2014-167",
    doi = "10.1103/PhysRevLett.114.221602",
    journal = "Phys. Rev. Lett.",
    volume = "114",
    number = "22",
    pages = "221602",
    year = "2015"
}

@article{Bogers:2018zeg,
    author = "Bogers, Mark P. and Brauner, Tom\'a\v{s}",
    title = "{Lie-algebraic classification of effective theories with enhanced soft limits}",
    eprint = "1803.05359",
    archivePrefix = "arXiv",
    primaryClass = "hep-th",
    doi = "10.1007/JHEP05(2018)076",
    journal = "JHEP",
    volume = "05",
    pages = "076",
    year = "2018"
}

@article{Joung:2015jza,
    author = "Joung, Euihun and Mkrtchyan, Karapet",
    title = "{Partially-massless higher-spin algebras and their finite-dimensional truncations}",
    eprint = "1508.07332",
    archivePrefix = "arXiv",
    primaryClass = "hep-th",
    doi = "10.1007/JHEP01(2016)003",
    journal = "JHEP",
    volume = "01",
    pages = "003",
    year = "2016"
}

@article{Albrychiewicz:2020ruh,
    author = "Albrychiewicz, Emil and Neiman, Yasha",
    title = "{Scattering in the static patch of de Sitter space}",
    eprint = "2012.13584",
    archivePrefix = "arXiv",
    primaryClass = "hep-th",
    doi = "10.1103/PhysRevD.103.065014",
    journal = "Phys. Rev. D",
    volume = "103",
    number = "6",
    pages = "065014",
    year = "2021"
}

@article{Marolf:2012kh,
    author = "Marolf, Donald and Morrison, Ian A. and Srednicki, Mark",
    title = "{Perturbative S-matrix for massive scalar fields in global de Sitter space}",
    eprint = "1209.6039",
    archivePrefix = "arXiv",
    primaryClass = "hep-th",
    doi = "10.1088/0264-9381/30/15/155023",
    journal = "Class. Quant. Grav.",
    volume = "30",
    pages = "155023",
    year = "2013"
}

@article{Melville:2024ove,
    author = "Melville, Scott and Pimentel, Guilherme L.",
    title = "{A de Sitter S-matrix from amputated cosmological correlators}",
    eprint = "2404.05712",
    archivePrefix = "arXiv",
    primaryClass = "hep-th",
    doi = "10.1007/JHEP08(2024)211",
    journal = "JHEP",
    volume = "08",
    pages = "211",
    year = "2024"
}

@article{BBBB,
    author = "Pajer, Enrico",
    title = "{Building a Boostless Bootstrap for the Bispectrum}",
    eprint = "2010.12818",
    archivePrefix = "arXiv",
    primaryClass = "hep-th",
    doi = "10.1088/1475-7516/2021/01/023",
    journal = "JCAP",
    volume = "01",
    pages = "023",
    year = "2021"
}

@article{Gazeau:2010mn,
    author = "Gazeau, Jean-Pierre and Siegl, Petr and Youssef, Ahmed",
    title = "{Krein Spaces in de Sitter Quantum Theories}",
    eprint = "1001.4810",
    archivePrefix = "arXiv",
    primaryClass = "hep-th",
    doi = "10.3842/SIGMA.2010.011",
    journal = "SIGMA",
    volume = "6",
    pages = "011",
    year = "2010"
}

@article{Sun:2021thf,
    author = "Sun, Zimo",
    title = "{A note on the representations of SO(1,d + 1)}",
    eprint = "2111.04591",
    archivePrefix = "arXiv",
    primaryClass = "hep-th",
    doi = "10.1142/S0129055X24300073",
    journal = "Rev. Math. Phys.",
    volume = "37",
    number = "01",
    pages = "2430007",
    year = "2025"
}

@article{Goon:2011qf,
    author = "Goon, Garrett and Hinterbichler, Kurt and Trodden, Mark",
    title = "{Symmetries for Galileons and DBI scalars on curved space}",
    eprint = "1103.5745",
    archivePrefix = "arXiv",
    primaryClass = "hep-th",
    doi = "10.1088/1475-7516/2011/07/017",
    journal = "JCAP",
    volume = "07",
    pages = "017",
    year = "2011"
}

@article{Du:2025glv,
    author = "Du, Zong-Zhe",
    title = "{Soft Unification of Exceptional EFTs in de Sitter space}",
    eprint = "2509.08979",
    archivePrefix = "arXiv",
    primaryClass = "hep-th",
    month = "9",
    year = "2025"
}

@article{Basile:2024ydc,
    author = "Basile, Thomas and Joung, Euihun and Mkrtchyan, Karapet and Mojaza, Matin",
    title = "{Spinor-helicity representations of particles of any mass in dS4 and AdS4 spacetimes}",
    eprint = "2401.02007",
    archivePrefix = "arXiv",
    primaryClass = "hep-th",
    reportNumber = "Imperial-TP-KM-2024-01",
    doi = "10.1103/PhysRevD.109.125003",
    journal = "Phys. Rev. D",
    volume = "109",
    number = "12",
    pages = "125003",
    year = "2024"
}

@article{Goon:2011uw,
    author = "Goon, Garrett and Hinterbichler, Kurt and Trodden, Mark",
    title = "{A New Class of Effective Field Theories from Embedded Branes}",
    eprint = "1103.6029",
    archivePrefix = "arXiv",
    primaryClass = "hep-th",
    doi = "10.1103/PhysRevLett.106.231102",
    journal = "Phys. Rev. Lett.",
    volume = "106",
    pages = "231102",
    year = "2011"
}

@article{Maldacena:2011nz,
    author = "Maldacena, Juan M. and Pimentel, Guilherme L.",
    archivePrefix = "arXiv",
    doi = "10.1007/JHEP09(2011)045",
    eprint = "1104.2846",
    journal = "JHEP",
    pages = "045",
    primaryClass = "hep-th",
    reportNumber = "PUPT-2371",
    title = "{On graviton non-Gaussianities during inflation}",
    volume = "09",
    year = "2011"
}

@article{Conjecture,
      author         = "Baumann, Daniel and Green, Daniel and Lee, Hayden and
                        Porto, Rafael A.",
      title          = "{Signs of Analyticity in Single-Field Inflation}",
      journal        = "Phys. Rev.",
      volume         = "D93",
      year           = "2016",
      number         = "2",
      pages          = "023523",
      doi            = "10.1103/PhysRevD.93.023523",
      eprint         = "1502.07304",
      archivePrefix  = "arXiv",
      primaryClass   = "hep-th",
      reportNumber   = "ICTP-SAIFR-15-252",
      SLACcitation   = "%%CITATION = ARXIV:1502.07304;%%"
}

@article{Raju:2012zr,
    author = "Raju, Suvrat",
    title = "{New Recursion Relations and a Flat Space Limit for AdS/CFT Correlators}",
    eprint = "1201.6449",
    archivePrefix = "arXiv",
    primaryClass = "hep-th",
    reportNumber = "HRI-ST-1201",
    doi = "10.1103/PhysRevD.85.126009",
    journal = "Phys. Rev. D",
    volume = "85",
    pages = "126009",
    year = "2012"
}

@article{Bonifacio:2018zex,
    author = "Bonifacio, James and Hinterbichler, Kurt and Joyce, Austin and Rosen, Rachel A.",
    title = "{Shift Symmetries in (Anti) de Sitter Space}",
    eprint = "1812.08167",
    archivePrefix = "arXiv",
    primaryClass = "hep-th",
    doi = "10.1007/JHEP02(2019)178",
    journal = "JHEP",
    volume = "02",
    pages = "178",
    year = "2019"
}

@article{Bonifacio:2021mrf,
    author = "Bonifacio, James and Hinterbichler, Kurt and Joyce, Austin and Roest, Diederik",
    title = "{Exceptional scalar theories in de Sitter space}",
    eprint = "2112.12151",
    archivePrefix = "arXiv",
    primaryClass = "hep-th",
    doi = "10.1007/JHEP04(2022)128",
    journal = "JHEP",
    volume = "04",
    pages = "128",
    year = "2022"
}

@article{Cheung:2015ota,
    author = "Cheung, Clifford and Kampf, Karol and Novotny, Jiri and Shen, Chia-Hsien and Trnka, Jaroslav",
    title = "{On-Shell Recursion Relations for Effective Field Theories}",
    eprint = "1509.03309",
    archivePrefix = "arXiv",
    primaryClass = "hep-th",
    reportNumber = "CALT-TH-2015-047",
    doi = "10.1103/PhysRevLett.116.041601",
    journal = "Phys. Rev. Lett.",
    volume = "116",
    number = "4",
    pages = "041601",
    year = "2016"
}

\end{document}